\documentstyle[epsfig,12pt]{article}

\advance \topmargin by -\headheight
\advance \topmargin by -\headsep
     
\evensidemargin \oddsidemargin
 \date{}

\def\nonu{\nonumber}

\def\br{\begin{eqnarray}}
\def\er{\end{eqnarray}}
\def\be{\begin{equation}}
\def\ee{\end{equation}}
\def\0{\nonumber}

\def\({\left(}
\def\){\right)}

\relax

\def\a{\alpha}
\def\b{\beta}

\def\d{\delta} 

\def\eps{\epsilon}

\def\l{\lambda}

\def\o{\over}

\def\pa{\partial}
\def\pr{\prime}

\def\lie{{\cal G}}

\def\rlx{\relax\leavevmode}
\def\inbar{\vrule height1.5ex width.4pt depth0pt}
\def\IZ{\rlx\hbox{\sf Z\kern-.4em Z}}
\def\IR{\rlx\hbox{\rm I\kern-.18em R}}
\def\IC{\rlx\hbox{\,$\inbar\kern-.3em{\rm C}$}}
\def\one{\hbox{{1}\kern-.25em\hbox{l}}}

\begin{document}
\setlength{\baselineskip}{5mm}

\noindent{\large\bf
 Classical Integrability of  Non Abelian Affine Toda Models\footnote{talk given at the
  XXIII International Colloquium on Group Theoretical Methods in
Physics, Dubna, August 2000} 
}\vspace{4mm}

\noindent{
{J.F. Gomes}$^a$,
E. P. Gueuvoghlanian$^a$,
G.M. Sotkov$^a$
and
 A.H. Zimerman$^a$
}\vspace{1mm}

\noindent{\small
$^a$Instituto de F\'\i sica Te\'orica - IFT/UNESP, 
Rua Pamplona 145,01405-900, S\~ao Paulo - SP, Brazil
}\vspace{4mm}

 A class of non abelian affine Toda models is
 constructed in terms of the axial and vector gauged WZW model.  
 It is shown that the multivacua structure of the potential together 
 with non abelian nature of the zero grade subalgebra
 allows soliton solutions with non trivial electric and topological charges.  
 Their zero curvature representation and the classical $r$-matrix are also constructed 
 in order to prove their classical integrability.

\section{Introduction}

The abelian affine Toda field theories provide a large class of integrable models in two dimensions
associated to an affine Lie algebra $\tilde \lie$ (loop algebra) admiting  solitons solutions.  The
Toda fields are defined to parametrize a finite dimensional  abelian manifold (Cartan subalgebra of  
$\tilde \lie$) and their solitonic character is a consequence of the infinite dimensional Lie
algebraic structure responsible for the multivacua configuration leading to a nontrivial topological
structure.

A more general class of affine Toda models is obtained by introducing a non abelian structure to the
abelian manifold (Cartan subalgebra of  
$\tilde \lie$) parametrized by the Toda fields. 
 A systematic  manner in classifying the Toda models
\cite{lez-sav}
is in terms of a grading operator $Q$ that decomposes the affine lie algebra $\tilde \lie = \oplus
\lie_i$, where the graded subspaces are defined by $[Q, \lie_i ] = i \lie_i$.  The Toda fields are
defined to parametrize the zero grade subspace $\lie_0 \subset \lie$.  

In this note we discuss a systematic construction  of the simplest affine non abelian Toda models
associated to $\lie_0 = SL(2) \otimes U(1)^{r-1}$ defined by the gradation  $ Q= h^{\pr} \hat d +
\sum_{i\neq a} {{2 \l_i \cdot H}\o {\a_i^2}}$, where $\hat d, \l_i$, $\a_i$ are the derivation
operator, fundamental weights and simple roots respectively  and $h^{\pr}$ is a real number
specified latter.  More general and interesting structures arises when the Toda fields lie in the
coset $\lie_0 /\lie_0^0$ \cite{ime} where $\lie_0^0 \in \lie_0$
 is some specific element of $\lie_0$.  The simplest model associated to $\lie_0 / \lie_0^0 =
 SL(2)/ U(1)$, is the Lund-Regge (or complex sine-Gordon) model \cite{lund}.
 
 In section 2 we construct the general action for the affine non abelian Toda models associated to
 the coset ${{SL(2) \otimes U(1)^{r-1}}\o {U(1)}}$ in terms of the
 gauged two-loop Wess-Zumino-Witten (WZW) model and explicitly show that the non abelian structure of
 $\lie_0$ allows a global gauge invariance   responsible for the conservation of an electric charge.
 In general, the models  admit electric (Noether) and magnetic (topological ) charges and 
 for imaginary coupling constant $\b $ ($\b^2 = -{{2\pi}\o {k}}$) they admit electrically 
 charged topological solitons constructed in
 \cite{emilio}.  It was
 shown in  {\cite{dual} that there are two inequivalent ways to gauge fix the 
  $\lie_0^0$ degree of freedom, {\it axial}
 or {\it vector}, leading to a pair of  actions T-dual to  each other.

  In section 3 we construct systematically the zero curvature representation for the affine non
  abelian Toda models, showing therefore, the existence of infinite number of conserved charges.  We
  next construct the classical $r$-matrix and derive the fundamental Poisson bracket relation 
  which ensures the involution of the conserved charges.

\section{Construction of the Model}
The generic NA Toda models  are classified
 according to a $\lie_0 \subset \lie$ embedding  induced
by the grading operator $Q$ decomposing an finite or infinite Lie algebra 
$\lie = \oplus _{i} \lie _i $ where $
[Q,\lie_{i}]=i\lie_{i}$ and $ [\lie_{i},\lie_{j}]\subset \lie_{i+j}$.  A group
element $g$ can then be written in terms of the Gauss decomposition as 
\be
g= NBM
\label{1}
\ee
where $N=\exp \lie_< $, $B=\exp \lie_{0} $ and
$M=\exp \lie_> $.  The physical fields lie in the zero grade subgroup $B$ 
and the
models we seek correspond to the coset $H_- \backslash G/H_+ $, for $H_{\pm} $ generated by
positive/negative  grade operators.

For consistency with the hamiltonian reduction formalism, the phase space of
the G-invariant WZNW model is  reduced by specifying the constant
generators $\eps_{\pm}$ of grade $\pm 1$.  In order to derive 
 an action for $B  $,  invariant under 
\begin{eqnarray}
g\longrightarrow g^{\prime}=\alpha_{-}g\alpha_{+},
\label{2}
\end{eqnarray}
where $\a_{\pm}(z, \bar z)$ lie in the positive/negative grade subgroup
 we have to introduce a set of  auxiliary
gauge fields $A \in \lie _{<} $ and $\bar A \in \lie _{>}$ transforming as 
\begin{eqnarray}
A\longrightarrow A^{\prime}=\alpha_{-}A\alpha_{-}^{-1}
+\alpha_{-}\partial \alpha_{-}^{-1},
\quad \quad 
\bar{A}\longrightarrow \bar{A}^{\prime}=\alpha_{+}^{-1}\bar{A}\alpha_{+}
+\bar{\partial}\alpha_{+}^{-1}\alpha_{+}.
\label{3}
\end{eqnarray}
The resulting action is the $G/H (= H_- \backslash G/H_+ )$
 gauged WZNW    
\begin{eqnarray}
S_{G/H}(g,A,\bar{A})&=&S_{WZNW}(g)
\nonumber
\\
&-&\frac{k}{2\pi}\int d^2x Tr\( A(\bar{\partial}gg^{-1}-\epsilon_{+})
+\bar{A}(g^{-1}\partial g-\epsilon_{-})+Ag\bar{A}g^{-1}\) .
\nonumber
\end{eqnarray}
Since the action $S_{G/H}$ is $H$-invariant,
 we may choose $\alpha_{-}=N_{}^{-1}$
and $\alpha_{+}=M_{}^{-1}$. From the orthogonality  of the graded 
subpaces, i.e. $Tr( \lie _i\lie _j ) =0, i+j \neq 0$, we find
\begin{eqnarray}
S_{G/H}(g,A,\bar{A})&=&S_{G/H}(B,A^{\prime},\bar{A}^{\prime})
\nonumber
\\
&=&S_{WZNW}(B)-\frac{k}{2\pi}
\int d^2x Tr[-A^{\prime}\epsilon_{+}-\bar{A}^{\prime}\epsilon_{-}
+A^{\prime}B\bar{A}^{\prime}B^{-1}],
\label{14}
\end{eqnarray}
where 
\begin{eqnarray}
S_{WZNW}=- \frac{k}{4\pi }\int d^2xTr(g^{-1}\partial gg^{-1}\bar{\partial }g)
+\frac{k}{24\pi }\int_{D}\epsilon^{ijk}
Tr(g^{-1}\partial_{i}gg^{-1}\partial_{j}gg^{-1}\partial_{k}g)d^3x,
\label{3a}
\end{eqnarray}
and the topological term denotes a surface integral  over a ball $D$
identified as  space-time.


Performing the integration  over the auxiliary fields $A$ and $\bar A$, 
the functional integral 
\be
 Z_{\pm}=\int DAD\bar{A}\exp (F_{\pm}),
\label{fpm}
\ee 
 where 
 \be
F_{\pm} = {-{k\o {2\pi}}}\int \(Tr
 (A - B {\eps_-} B^{-1})B(\bar A - B^{-1} {\eps_+} B) B^{-1}\)d^2x
\label{fmm}
\ee
yields the effective action
\be
S = S_{WZNW} (B) + {{k\o {2\pi}}} \int Tr \( \eps_+ B  \eps_- B^{-1}\)d^2x
\label{spm}
\ee
The action (\ref{spm}) describe integrable perturbations of the $\lie_0$-WZNW model. 
 Those perturbations are classified in
terms of the possible constant grade $\pm 1$ operators $\eps _{\pm}$.

 More interesting cases 
arises in
connection with non abelian embeddings $\lie_0 \subset \lie $.  In particular, if we 
supress one
of the fundamental weights from $Q$, the zero grade subspace $\lie_0$,
acquires a nonabelian 
structure
$sl(2)\otimes u(1)^{rank \lie -1}$.  Let us consider for instance 
$Q= h^{\pr} d + \sum_{i\neq a}^{r}\frac{2\lambda_{i}\cdot H}{\alpha_{i}^{2}}$, 
where $h^{\pr} =0$ or
$h^{\pr} \neq 0$ corresponding to
 the Conformal or Affine  nonabelian (NA) Toda 
respectively.
The absence of $\lambda_a$ in $Q$
 prevents the contribution of the simple root step operator
$E_{\a_a}^{(0)}$ in constructing $\eps_+$. It in fact, allows for 
reducing the phase space 
even further.  This fact can be understood by enforcing the 
nonlocal constraint $J_{Y \cdot H} = \bar J_{Y \cdot H} = 0$
where $Y$ is such that $[Y\cdot H , \eps_{\pm}] = 0$ and 
$J=g^{-1}\partial g$ and $\bar{J}=-\bar{\partial}gg^{-1}$.  Those generators 
of $\lie_0$ commuting with
$\eps_{\pm}$ define a subalgebra $\lie_0^0 \subset \lie_0 $.
 Such subsidiary constraint is incorporated into the action by
  requiring symmetry under \cite{plb} 
\begin{eqnarray}
g\longrightarrow g^{\prime}=\alpha_{0}g\alpha_{0}'
\label{5}
\end{eqnarray} 
where we shall consider   $\a^{\pr}_{0}
 =\alpha_{0}(z, \bar z) \in \lie_0^0 $, i.e., {\it axial symmetry} (the 
 {\it vector }
 gauging is obtained by choosing $\a^{\pr}_{0}
 ={\alpha_{0}}^{-1}(z, \bar z) \in \lie_0^0 $). 
Auxiliary gauge fields $A_0 = a_0 Y\cdot H$ and $\bar A_0= 
\bar a_0 Y\cdot H\in  \lie_{0}^{0}$  are 
introduced to construct  an invariant action under transformations  (\ref{5}) 
\br 
S(B,{A}_{0},\bar{A}_{0} ) &=& S(g_0^f,{A^{\pr}}_{0},\bar{A^{\pr}}_{0} )  
 = S_{WZNW}(B)+ 
 {{k\o {2\pi}}} \int Tr \( \eps_+ B  \eps_- B^{-1}\) d^2x\nonu \\   
  &-&{{k\o {2\pi}}}\int Tr\(  A_{0}\bar{\partial}B
B^{-1} + \bar{A}_{0}B^{-1}\partial B
+ A_{0}B\bar{A}_{0}B^{-1} + A_{0}\bar{A}_{0} \)d^2x \nonu \\
\label{aa}
\er
where the auxiliary fields transform as
\begin{eqnarray}
A_{0}\longrightarrow A_{0}^{\prime}=A_{0}-\alpha_{0}^{-1}\partial \alpha_{0},
\quad \quad 
\bar{A}_{0}\longrightarrow \bar{A}_{0}^{\prime}=\bar{A}_{0}
- \bar{\partial}\alpha_{0}^{\pr}(\alpha_{0}^{\pr})^{-1}.
\nonu 
\end{eqnarray}
 Such residual gauge symmetry allows us to eliminate an  extra field
 associated to $Y\cdot H$.
  Notice that the physical fields
 $g_0^f$ lie in the coset $\lie_0 /{\lie_0^0} = ({{sl(2)\otimes
u(1)^{rank \lie -1}})/u(1)}$ of dimension $rank
\lie +1$ and are classified according to the 
gradation $Q$.  It therefore follows that $S(B,A_0,\bar{A_0})=
 S(g_0^f,A_0^{\pr},\bar{A_0}^{\pr})$.

In \cite{ime} a detailed study of the gauged WZNW construction 
for  finite dimensional Lie algebras leading to Conformal NA Toda
models was presented.  The study of its symmetries was given in ref. \cite{plb}. 
 Here we generalize the construction of ref. \cite{ime} to
infinite dimensional Lie algebras leading to NA Affine 
Toda models characterized 
by the broken
conformal symmetry and by the  presence of
solitons. 
 
Consider the Kac-Moody
algebra ${\widehat \lie }$
\br
[T_m^a ,T_n^b]  =  f^{abc} T^c_{m+n} + {\hat c}m \d_{m+n} \d^{ab} \nonu 
\er
\br
[{\hat d} , T^a_n] = nT^a_n ;\quad [{\hat c}, T^a_n] = [{\hat c},{\hat d} ] = 0
\label{km}
\er

The NA Toda models we shall be constructing are associated to 
gradations of the type
$Q_a(h^{\pr}) = h^{\pr}_a d + 
\sum_{i\neq a}^{r}\frac{2\lambda_{i}\cdot H^{(0)}}{\alpha_{i}^{2}}$, 
where $h^{\pr}_a$ is chosen
such that the gradation,
 $Q_a(h^{\pr})$, acting on
infinite dimensional Lie algebra $ \hat \lie$ ensures that the zero grade 
subgroup $\lie_0$ coincides with its
counterpart obtained with $Q_a(h^{\pr}=0)$ 
 acting on the Lie algebra $\lie $ of finite dimension apart from two  comuting 
 generators $\hat {c}$ and $\hat {d}$.  Since they commute with $\lie_0$,
  the kinetic part
 decouples such that the conformal and the affine singular NA-Toda  models 
 differ only by the
 potential term characterized by $\hat \eps_{\pm}$. 

The integration over the auxiliary gauge fields $A$ and $\bar A$ require 
explicit
parametrization of $B$. 
\begin{eqnarray}
B=\exp (\tilde {\chi} E_{-\alpha_{a}}^{(0)})
 \exp (   R \sum_{i=1}^{r}{{Y_i}} {{H_i^{(0)}}}+\Phi (H)+ \nu \hat {c} +
  \eta \hat {d})\exp (\tilde {\psi} E_{\alpha_{a}}^{(0)})
 \label{63}
 \end{eqnarray}
where $ \Phi (H) 
=\sum_{j=1}^{r}\sum_{i=2}^{r}\varphi_{i}{{X}}_i^j {{H_j}^{(0)}}$,
 where $\sum_{j=1}^r{{Y_j}}  {{X^j_i}} =0, i=2, \cdots ,r$.
After gauging away the nonlocal field $R$, the factor group element becomes    
\be
g_0^f=\exp (\chi E_{-\alpha_{a}}^{(0)})
 \exp (   \Phi (H)+ \nu \hat {c} + \eta \hat {d})\exp (\psi E_{\alpha_{a}}^{(0)})
 \label{63a}
\ee
where $\chi = \tilde {\chi}e^{{1\o 2}(Y\cdot \a_a)R}, \quad 
\psi = \tilde {\psi}e^{{1\o 2}(Y\cdot \a_a)R}$
We therefore get for the zero grade component
\br
F_0 &=&{-{k\o {2\pi}}}\int Tr\(  A_{0}\bar{\partial}g_0^f
(g_0^f)^{-1} + \bar{A}_{0}(g_0^f)^{-1}\partial g_0^f
+ A_{0} g_0^f\bar{A}_{0}(g_0^f)^{-1} + A_{0}\bar{A}_{0} \)d^2x \nonu \\
&=&{-{k\o {2\pi}}}\int \( a_0 \bar a_0 2Y^2\Delta -  2({{\a_a \cdot Y}\o
{\a_a^2}})(\bar a_0\psi \pa \chi + a_0 \chi \bar \pa \psi )e^{\Phi (\a_a)}\)
d^2x
\label{del}
\er
where $\Delta = 1 + {{(Y \cdot \a_a )^2}\o {2Y^2}}\psi \chi e^{\Phi (\a_a )}$,
$[\Phi (H), E_{\alpha_{a}}^{(0)}] = \Phi (\a_a)E_{\alpha_{a}}^{(0)}$. 

The effective action is obtained by integrating over the auxiliary 
fields $A_0, \bar A_0$, 
\be
Z_0 = \int DA_{0}D\bar{A}_{0}\exp (F_{0}) 
\ee  
 The total action (\ref{aa}) is therefore given as
 \be
 S= -{k \o {4\pi}}\int \( Tr (\pa \Phi(H)\bar \pa \Phi(H)) + 
 {{2\bar \pa \psi \pa \chi }\o \Delta
 }e^{\Phi(\a_a)} + \pa \eta \bar \pa \nu +
 \pa \nu \bar \pa \eta 
    - 2 Tr (\hat {\eps_+} g_0^f\hat {\eps_-} (g_0^f)^{-1}) \)d^2x
 \label{action}
 \ee
Note that the second term in (\ref{action}) contains both symmetric and
antisymmetric parts:
\begin{eqnarray}
\frac{e^{\Phi(\a_a)} } {\Delta}\bar{\partial}\psi \partial \chi
={1\o 4}\frac{ e^{\Phi(\a_a)} }
{\Delta}(g^{\mu \nu}\partial_{\mu}\psi\partial_{\nu}\chi
+\epsilon^{\mu \nu}\partial_{\mu}\psi \partial_{\nu}\chi ),
\end{eqnarray}
where $g_{\mu \nu}$ is the 2-D metric of signature $ g_{\mu
\nu}= diag (1,-1)$, $z = t + x\;\; \bar z = t -x$.
 For $n=1$ ($\lie \equiv A_{1}$, $\Phi (\a_1)$ is zero) the
antisymmetric term is a total derivative:
\begin{eqnarray}
\epsilon^{\mu \nu}\frac{\partial_{\mu}\psi \partial_{\nu}\chi}{1+\psi \chi}
=\frac{1}{2}\epsilon^{\mu \nu}\partial_{\mu}
\left( \ln \left\{ 1+\psi \chi \right\}
\partial_{\nu}\ln{\frac{\chi}{\psi}}\right),
\end{eqnarray}
and it can be neglected.  This $A_{1}$-NA-Toda model (in the conformal case), 
is known to describe the
2-D black hole solution for (2-D) string theory \cite{Witten1}.
The
$\lie $-NA conformal Toda model can be used in the
description of specific (n+1)-dimensional black string theories 
\cite{gervais-saveliev},
 with  n-1-flat and
2-non flat directions ($g^{\mu
\nu}G_{ab}(X)\partial_{\mu}X^{a}\partial_{\nu}X^{b}$, $X^{a}=(\psi ,\chi
,\varphi_{i})$), containing axions ($\epsilon^{\mu
\nu}B_{ab}(X)\partial_{\mu}X^{a}\partial_{\nu}X^{b}$) and tachyons
($\exp \left\{ -k_{ij}\varphi_{j}\right\} $), as well.  The affine 
$ A_{1}$-NA Toda theory with $\eps _{\pm} = H^{(\pm )}$
correspond to the Lund-Regge model describing charged solitons.

It is clear that the presence of the $ e^{\Phi(\a_a)}$ in (\ref{action})
is responsible for the
 antisymmetric  tensor
generating CPT breaking terms. 
On the other hand, notice that $\Phi(\a_a)$ depend upon the subsidiary nonlocal
constraint $J_{Y \cdot H} = \bar J_{Y \cdot H} = 0$ and hence upon the choice of
the vector {{Y}}.  It is defined to be orthogonal to all roots contained in
$\eps_{\pm}$.  A Lie algebraic condition for the absence of axionic terms was found in \cite{ime} and
has provided a construction of a family of torsionless NA Toda models in 
\cite{dual}.  All vector models are CPT invariant by construction.

The action (\ref{action}) is invariant under the global $U(1)$ transformation 
\begin{equation}
{\psi }\rightarrow e^{i\epsilon }{\psi }, \quad \quad 
{\chi }\rightarrow e^{-i\epsilon }{\chi }
 \label{13.84}
\end{equation}
The corresponding Noether current is 
\begin{equation}
J^{\mu }=-\frac{ik}{8\pi }\frac{e^{\Phi(\a_a)}}{\Delta }\{{\psi 
}\left( g^{\nu \mu }\partial _{\nu }{\chi }-\epsilon ^{\nu \mu
}\partial _{\nu }{\chi }\right) -{\chi }\left( g^{\nu
\mu }\partial _{\nu }{\psi }+\epsilon ^{\nu \mu }\partial _{\nu }
{\psi }\right) \}
\end{equation}
and the electric charge is given in terms of the nonlocal field $R$ defined 
below in
(\ref{bh}) as
\begin{equation}
Q_{}=\int J_o dx= -\frac{ik}{4\pi }\left( \frac{r}{r+1}\right) \left[
R(x \rightarrow \infty )-R(x \rightarrow -\infty )\right].  \label{13.87}
\end{equation}

Apart from the global $U(1)$   symmetry (\ref{13.84}) there is a discrete 
set of field transformations leaving the action (\ref{action}) unchanged.
Such transformations  (for imaginary $\b $, $\b \rightarrow i \b_0$) 
 give rise to multivacua configuration and 
hence to nontrivial topological charges
\br
 Q_j &=& \int_{-\infty}^{+\infty} J^0_{j} dx,  \quad J_j^{\mu} = 
 -i {{2r}\o { \b}} \eps^{\mu \nu } \pa_{\nu} \varphi_j, \quad j=2,
 \cdots r \nonu \\
 Q_{\theta} &=& \int_{-\infty}^{+\infty} J_{\theta}^{0} dx    
  \quad J_{\theta}^{\mu} = -i {{1}\o { 2\b^2}} \eps^{\mu \nu } \pa_{\nu} 
 ln \( {{\chi }\o {\psi}}\)
 \label{topch}
 \er
Let us explicitly  
consider the $A_{r}^{(1)}$ model described by the 
Lagrangean density (\ref{lagran}) (with fields rescaled by $\varphi_i \rightarrow \b \varphi_i, 
\chi \rightarrow \b \chi, \psi \rightarrow \b \psi$, $\b^2 = -{{2\pi}\o {k}}$)
invariant under the following set of
discrete transformations, 
\br
\varphi_{j}^{\pr} = \varphi_{j} + {{2\pi (j-1) N}\o {\b_0 r}}, j=2, \cdots r,\quad 
\chi^{\pr} = e^{i\pi ({{N\o r} + s_2}) }\chi, \quad \psi^{\pr} = e^{i\pi ({{N\o r} + s_1}) }\psi
\er
where $s_1, s_2$ are both even or  odd integers and the following CP transformations (P: $x \rightarrow -x$)
\br
\varphi_{j}^{\pr \pr} = \varphi_{j}, j=2, \cdots r,\quad
\chi^{\pr \pr} = \psi, \quad \psi^{\pr \pr} =\chi
\er
 
The minimum of the potential ( for the choice $\eta =0$)
 corresponds to the following field configuration 
\br
\varphi_{j}^{(N)} ={{2\pi (j-1) N}\o {\b_0 r}}, \quad \theta^{(L)} ={{1}\o {2i \b_0}}ln \( {{\chi }\o {\psi }}\) = 
 {{\pi L}\o {2\b_0}}, \quad \rho^{(0)} = 0 \quad j=2, \cdots r
 \label{vac}
\er
where $N, L$ are arbitrary integers, and the new fields $\theta$ and $\rho$ are defined as
\be
\psi = {1\o {\b_0}} e^{i \b_0 ({1\o 2} \varphi_2 - \theta )}\sinh (\b_0 \rho ), \quad 
\chi = {1\o {\b_0}} e^{i \b_0 ({1\o 2} \varphi_2 + \theta )}\sinh (\b_0 \rho )
\ee
In fact eqns, (\ref{vac}) also represent constant solutions of the eqns.
 of motion (\ref{13.20})-(\ref{1333}) which allows us to derive
the values of the topological charge (\ref{topch}): $ Q_j = {{4 \pi }\o {\b_0^2}}(j-1)(N_+-N_-),
 \;\; Q_{\theta} = {{2 \pi}\o {\b_0^2}} (L_+ - L_-)$

 
\section{Zero Curvature and Equations of Motion}

The equations of motion for the NA Toda models are known to be of the form
\cite{lez-sav}
\be 
\bar \pa (B^{-1} \pa B) + [\hat {\eps_-}, B^{-1} \hat {\eps_+} B] =0, 
\quad \pa (\bar \pa B B^{-1} ) - [\hat {\eps_+}, B\hat {\eps_-} B^{-1}] =0
\label{eqmotion}
\ee
 The subsidiary constraint $J_{Y \cdot H^{(0)}} =
  Tr(B^{-1} \pa B Y\cdot H^{(0)})$ and $
  \bar J_{Y \cdot H^{(0)}} = 
  Tr(\bar \pa B B^{-1}Y \cdot H^{(0)} )=        0$ can be
 consistently imposed  since $[Y\cdot H^{(0)}, \hat {\eps_{\pm}}]=0$ as can be 
 obtained from
 (\ref{eqmotion}) by taking the trace with $Y.H^{(0)}$.  Solving those equations 
 for the nonlocal
 field $R$ yields, 
 \be
 \pa R = ({{Y\cdot \a_a}\o {Y^2}}) {{\psi \pa \chi }\o \Delta }e^{\Phi(\a_a)},
 \;\; \;\;\; 
\bar \pa R = ({{Y\cdot \a_a}\o {Y^2}}) {{\chi \bar \pa \psi }\o \Delta }e^{\Phi(\a_a)}
\label{bh}
\ee
The equations of motion for the fields $\psi, \chi $ and $\varphi_i, i=2,
\cdots , r$  obtained from (\ref{eqmotion}) after imposing the subsidiary 
constraints
(\ref{bh}) coincide precisely with the Euler-Lagrange equations derived from
 (\ref{action}).  Alternatively, (\ref{eqmotion}) admits a
zero curvature representation $\pa \bar A - \bar \pa A + [A, \bar A] =0$ where
\be
A= B \hat {\eps_-}  B^{-1} ,\quad  
\bar A= -\hat {\eps_+}   - \bar \pa B B^{-1} 
\label{zcc}
\ee
Whenever the constraints (\ref{bh}) are incorporated into $A$ and $\bar A$ in
(\ref{zcc}), equations (\ref{eqmotion}) yields the zero curvature
representation of the NA singular Toda models. 

We shall be considering 
$\hat \lie =  A_r^{(1)}$, $ Q= r\hat d + \sum_{i=2}^{r} 2 {{\lambda_i \cdot
H^{(0)}}\o {\a_i^2}}$,  ${\sum_{i=1}^{r} {Y_i H_i^{(0)}}} = 
2{{\lambda_1 \cdot H^{(0)}}\o {\a_1^2}}$,
 $ { \sum_{j=1}^{r}{X_i^j  H_j^{(0)}}} =
h_i^{(0)} = {{2 \a_i\cdot H^{(0)}}\o {\a_i^2}}$ and  $\hat {\eps_{\pm}} = \mu \(\sum_{i=2}^{r} E_{\pm \a_i}^{(0)} +
 E_{\mp (\a_2 + \cdots + \a_r)}^{(\pm 1)}\)$,  $\mu >0$.

Using the explicit parametrization of $B$ given in (\ref{63}),   we find, in
a systematic manner, the following form for $A$ and $\bar A$
 
\br
A  &=& \mu (  \sum _{i=2}^r e^{  -\sum _{j=2}^r K_{i,j}\varphi _{j}} E_{-\alpha
_{i}}^{(0)}-\chi e^{-{1\o 2}R-2\varphi _{2}+\varphi _{3}}E_{-\alpha _{1}-\alpha
_{2}}^{(0)}  \nonu \\
 &+&\psi e^{{1\o 2}R + \varphi_r -\eta} E_{\a_1 + \cdots \a_r}^{(-1)} + 
 (1+\psi \chi
e^{-\varphi_2})E_{\a_2 +\cdots + \a_r}^{(-1)} e^{\varphi_2 + \varphi_r -\eta
})
\label{a}
\er 
and 
\br
\overline{A} =  \mu \( -{{\sum _{i=2}^r }}E_{\alpha
_{i}}^{(0)}-E_{-\alpha _{2}-...-\alpha _{r}}^{(1)}\) -\( 
\overline{\partial }\chi -\chi 
\overline{\partial }\varphi _{2}+
({1\o {2\lambda_1^2}}-1){{\chi ^{2}\overline{\partial }\psi }\o \Delta }
e^{-\varphi _{2}}\) e^{-{1\o 2}R} E_{-\alpha _{1}}^{(0)}  \nonu 
\er
\be
 -   {{\overline{\partial }\psi }\o \Delta }
 e^{{1\o 2}R-\varphi _{2}}E_{\alpha _{1}}^{(0)}-
\overline{\partial }\nu \hat {c}-\overline{\partial }\eta \hat {d} -{
{\sum_{i=2}^r }}\overline{\partial }\varphi _{i}h_{i}^{(0)}-{{\chi 
\overline{\partial }\psi}\o \Delta} e^{-\varphi _{2}}{
{\sum _{j=2}^r}}\left( \frac{r+1-j}{r}\right) h_{j}^{(0)},
\label{13.14}
\ee
 leading to the following equations of motion 
\begin{equation}
\partial \overline{\partial }\eta =0, \quad 
 \partial \overline{\partial }\nu =\mu ^{2}e^{\varphi _{r}-\eta }(e^{\varphi
_{2}}+{\psi }{\chi }),
\label{13.20}
\end{equation}

\begin{equation}
\partial \left( \frac{e^{-\varphi _{2}}\overline{\partial }{\psi }
}{\Delta }\right) +\left( \frac{r+1}{2r}\right) \frac{{\psi }
e^{-2\varphi _{2}}\partial {\chi }\overline{\partial }{
\psi }}{\Delta ^{2}}+\mu ^{2}e^{\varphi _{r}-\eta }{\psi }=0,
\label{13.22}
\end{equation}

\begin{equation}
\overline{\partial }\left( \frac{e^{-\varphi _{2}}\partial {\chi }
}{\Delta }\right) +\left( \frac{r+1}{2r}\right) \frac{{\chi }%
e^{-2\varphi _{2}}\partial {\chi }\overline{\partial }{%
\psi }}{\Delta ^{2}}+\mu ^{2}e^{\varphi _{r}-\eta }{\chi }=0
\label{13.23}
\end{equation}
 
\br
\pa \bar \pa \varphi_i &+& ({{r+1-i}\o {r}}){{\pa \chi \bar \pa \psi
e^{-\varphi_2}}\o {\Delta^2}} \nonu \\ 
&+& \mu^2 e^{\varphi_2 +\varphi_r -\eta }
\(1+({{i-1}\o {r}})\psi \chi e^{-\varphi_2  }\)
 - \mu^2 e^{-\sum_{j=2}^{r}K_{ij}\varphi_j}  =0,  
 \label{1333}
 \er
$i=2, \cdots r$, where we  have normalized $\a^2 =2$.
%
%
%
%
%
%
%
%

\section{ Classical r-matrix}

Consider the effective action for the conformal affine  $A_r$-NA Toda model
specified in the previous section  by the action  
\br
S_{eff}&=&-\frac{k}{4\pi }\int d^{2}x\( {\sum _{i,j=2}^{r}}
K _{i,j}\partial \varphi _{i}\overline{\partial }\varphi
_{j}+\partial \nu \overline{\partial }\eta +
\partial \eta \overline{\partial 
}\nu    \right. \nonu \\
 &+& 
\left. \frac{2e^{-\varphi _{2}}\partial {\chi }\overline{\partial }
{\psi }}{\Delta }-2\mu ^{2}\( 
\sum _{i=2}^{r} e^{ {-\sum _{j=2}^{r} }K_{i,j}\varphi
_{j}} +e^{\varphi _{r}+\varphi _{2}-\eta }(1+{\psi }
{\chi }e^{-\varphi_2})\) \) .
\label{lagran}
\er
where $g^{00}=-g^{11}=1$ and $ K_{i,j} = Tr (h_i^{(0)}h_j^{(0)})   $.  The canonical momenta are 
\be
\Pi_{\varphi _{k}}=-\frac{k}{8\pi }K _{k,i}\partial _{t}\varphi _{i}, \;\; k
=2, \cdots , r \quad
\Pi_{\nu }=-\frac{k}{8\pi }\partial _{t}\eta, \quad  
\Pi _{\eta }=-\frac{k}{8\pi }\partial _{t}\nu,
  \label{13.51}
\ee
\be
\Pi_{\chi }=-\frac{k}{4\pi }\frac{e^{-\varphi _{2}}\overline{
\partial }{\psi }}{\Delta } \quad \quad 
\Pi_{\psi }=-\frac{k}{4\pi }\frac{e^{-\varphi _{2}}\partial 
{\chi }}{\Delta }.
 \label{13.53}
\ee
   

Let 
$A = {1\o 2 }(A_0 + A_1 )$ and $\bar A = {1\o 2} (A_0 - A_1 )$ given by eqns.
(\ref{a}) and (\ref{13.14}). 
Consider the gauge transformation 
\be
A^{\pr}_{\mu } = S A_{\mu } S^{-1} - \pa_{\mu} S S^{-1}
\nonu
\ee
where $S = S_3 S_2 S_1$ and

\begin{equation}
S_{1}=\exp \( -\frac{\varphi _{1}}{2} 2{{\lambda_1 \cdot H^{(0)}}\o {\a_1^2}}
- {{\nu }\o {2}}\hat c - {{\eta }\o {2}}\hat d -
{{\sum _{i=3}^{r} }}\frac{\varphi _{i}}{2}h_{i}^{(0)} \) ,
\end{equation}

\be
S_{2}=\exp (-{\chi }E_{-\alpha _{1}}^{(0)}), \quad   
S_{3}=\exp ( -\frac{\varphi _{2}}{2}h_{2}^{(0)})  \label{13.58}
\ee

yielding

\br
A_{x}^{\pr} &= &-\frac{4\pi }{k}\left( \Pi _{\eta }\hat c +\Pi _{\nu
}\hat d \right) +\mu \sum _ {i=2}^{r}
e^{ -{1\o 2}\sum _{j=2}^{r}{K}_{i,j}\varphi _{j}}
(E_{\a_i}^{(0)} + E_{-\a_i}^{(0)}) 
\nonu \\
&+& 
\mu e^{{1\o 2}(\varphi _{r}-\eta )}({\psi }E_{\alpha
_{1}+...+\alpha _{r}}^{(-1)}+{\chi }E_{-(\alpha _{1}+...+\alpha
_{r})}^{(1)})+\mu e^{{1\o 2}(\varphi _{2}+\varphi _{r}-
\eta )}(E_{\alpha _{2}+...+\alpha _{r}}^{(-1)}+E_{-(\alpha
_{2}+...+\alpha _{r})}^{(1)}) \nonumber \\
&- &\frac{4\pi }{k}
 e^{{1\o 2}
\varphi _{2}}(\Pi _{{\chi }}E_{\alpha
_{1}}^{(0)}+\Pi _{{\psi }}E_{-\alpha _{1}}^{(0)})-\frac{4\pi }{k}
{{\sum _{l,k=1}^{r-1}}}({\tilde K }^{-1})_{l,k}\Pi
_{\varphi _{k+1}}h_{l+1}^{(0)} \nonu\\
&-&{{2\pi }\o {k}}({{r+1}\o r})(\psi \Pi_{\psi}
 + \chi \Pi_{\chi}){{2\lambda_1 \cdot H^{(0)}}\o {\a_1^2}}
\er
where $\tilde K $ denotes the matrix defined by $K $ removing the first row
and first column.  
The fundamental Poisson bracket relation (FPR) 
\begin{equation}
\{ A_{x}^{\prime  }(y,t){\otimes }
A_{x}^{\prime  }(z,t)\}_{PB} =[ r,A_{x}^{\prime  }(y,t)\otimes I + 
I \otimes A_{x}^{\prime }(z,t) ]
\delta (y-z),  \label{13.675}
\end{equation}
 can then be verified, where the l.h.s. is evaluated using 
 the canonical comutation relations and $r$ denotes the classical 
 $r$-matrix 
\begin{equation}
r=-\frac{2\pi }{k}[C^{+}-C^{-}],  \label{13.68}
\end{equation}
where
\be
C^{+}={\sum _{m=1}^{\infty}}
{{\sum _{a,b=1}^{r}}}\frac{\alpha _{b}^{2}}{2}(K^{-1})_{a,b}\left(
h_{a}^{(m)}\otimes h_{b}^{(-m)}\right) +\frac{1}{2}
{
\sum _{\alpha >0}}\frac{\alpha ^{2}}{2}
\left( E_{\alpha }^{(0)}\otimes E_{-\alpha
}^{(0)}\right) + \nonumber
\ee
\begin{equation}
+{\sum _{m=1}^{\infty }}{\sum _{\alpha >0}}
\frac{\alpha ^{2}}{2}\left[ E_{\alpha }^{(m)}\otimes E_{-\alpha
}^{(-m)}+E_{-\alpha }^{(m)}\otimes E_{\alpha }^{(-m)}\right],  \label{13.69}
\end{equation} 
\begin{equation}
C^{-}=\sigma C^{+}, \;\; {\rm where } \quad \sigma (A_{1}\otimes B_{1})...(A_{n}\otimes B_{n})=(B_{1}\otimes
A_{1})...(B_{n}\otimes A_{n}).
 \label{13.70}
\end{equation}
and 
\be
C_0 =  \sum_{a,b=1}^{r} K_{ab}^{-1} h_a^{(0)} \otimes h_b^{(0)} + { \hat c}\otimes \hat d + \hat d \otimes { \hat c}.
\ee
In order to evaluate the  r.h.s. of eqn. (\ref{13.675}) we follow the  arguments
of ref. \cite{Aratyn}.  Let 
\be
{\cal C }= \sum_{m= -\infty}^{\infty} \sum_{a,b=1}^{r} K_{ab}^{-1} T_a^m \otimes T_b^{-m} + \hat c \otimes \hat d + \hat d \otimes \hat c
\ee
be the Casimir operator satisfying $
[ {\cal C }, 1 \otimes T + T \otimes  1] = 0 $
It then follows from direct calculation that
\br
[C_{\pm}, 1\otimes h_a^{(0)} + h_a^{(0)}\otimes 1] =  [C_{\pm}, 1\otimes \hat d + \hat d\otimes 1] = 
[C_{\pm}, 1\otimes \hat c + \hat c\otimes 1] = 0 \nonumber \\
\er
\br
[C_{0}, 1\otimes E_{\b}^{(n)} + E_{\b}^{(n)}\otimes 1]  &=& 
\( h_{\b}^{(0)} \otimes E_{\b}^{(n)} + E_{\b}^{(n)}\otimes h_{\b}^{(0)} \) + 
n \( \hat c \otimes E_{\b}^{(n)} + E_{\b}^{(n)}\otimes \hat c \)\nonu \\
&          =& - [ C_+ + C_- , 1\otimes E_{\b}^{(n)} + E_{\b}^{(n)}\otimes 1]
\label{rmatr}
\er
where $\b $ denotes an arbitrary simple root.
Using the fact that the sum of a positive (negative) simple root with a negative (positive) root 
is never a positive (negative) root
we obtain,
\br
[C_+, 1\otimes E_{\b}^{(n)} + E_{\b}^{(n)}\otimes 1]&=& - 
\( E_{\b}^{(n)}\otimes h_{\b}^{(0)}+ X_+ \otimes X_- \) \nonu
\er
\br
[C_-, 1\otimes E_{\b}^{(n)} + E_{\b}^{(n)}\otimes 1]&=& - 
\(  h_{\b}^{(0)} \otimes E_{\b}^{(n)} + X_- \otimes X_+ \)
\label{cplus}
\er
where $X_{\pm}$ denote terms with positive (negative) root step operators.  By adding these two equations and comparing with 
(\ref{rmatr}), we conclude that the last terms in the r.h.s. of (\ref{cplus}) vanish, and therefore
\br
[C_+ - C_-, 1\otimes E_{\b}^{(0)} + E_{\b}^{(0)} \otimes 1]=  \(  h_{\b}^{(0)} \otimes E_{\b}^{(0)}-
 E_{\b}^{(0)}\otimes h_{\b}^{(0)}\).
 \label{c1}
 \er
Analogously we find
\br
[C_+ - C_-, 1\otimes E_{\mp \psi}^{(\pm 1)} &+& E_{\mp \psi}^{(\pm 1)}\otimes 1] =  -\(  h_{\psi}^{(0)} \otimes E_{\mp \psi}^{(\pm 1)}-
 E_{\mp \psi}^{(\pm 1)}\otimes h_{\psi}^{(0)} \) \nonu \\
 &+ & \(  \hat c \otimes E_{\mp \psi}^{(\pm 1)}-
 E_{-\psi}^{(1)}\otimes \hat c \) , \quad h_{\psi} = \psi \cdot H
 \label{c2}
 \er
and 
\br
& &[C_+ - C_- , 1\otimes E_{\pm ( \a_2+ \cdots \a_r)}^{(\mp 1)} + E_{\pm (\a_2 + \cdots + \a_r )}^{(\mp 1)}\otimes 1]\nonu \\
&=& 
 -\sum_{a=2}^{r}  \( h_{a}^{(0)} \otimes E_{\pm (\a_2 + \cdots + \a_r) }^{(\mp 1)} -
 E_{\pm (\a_2 + \cdots \a_r) }^{(\mp 1)}\otimes h_{a}^{(0)}\)  \nonu \\
 &+&   \( \hat c \otimes E_{\pm (\a_2 + \cdots \a_r) }^{(\mp 1)}-
 E_{\pm (\a_2 + \cdots \a_r)}^{(\mp 1)}\otimes \hat c \) + 2\( E_{\pm \psi }^{(\mp 1)} \otimes E_{\mp \a_1}^{(0)} - 
 E_{\mp \a_1}^{(0)}\otimes E_{\pm \psi }^{(\mp 1)} \)
\label{c3} 
 \er
Using (\ref{c1})-(\ref{c3}) we find agreement for both sides of the fundamental Poisson bracket relation (\ref{13.675})
\section{Remarks on   Soliton Solutions }

A systematic and elegant method to obtain soliton solutions is to consider 
the dressing  of 
the gauge connection $ A$ and $\bar A$ in 
 (\ref{a}) and (\ref{13.14}) at  vacuum configuration \cite{luis},
i.e. $A_{vac}= \eps_-$ and  $\bar A_{vac} = -\eps_+ - \mu^2 z \hat c$.   A crucial ingredient  in
classifying the soliton solutions is the Heisenberg subalgebra generated by $\eps_{\pm}$ and its
eigenstates (vertex operators).  
We have considered in  \cite{emilio} the solutions for the 
$A_r^{(1)}$ model described in section 3 with vertices 
 constructed 
explicitly  in ref. \cite{vertex}.
 The  solitons were  classified according to the vertices and give
rise to neutral solutions (solutions of the $A_{r-1}$ affine abelian Toda
model with $\psi =\chi =0$)  and nontrivial charged solutions.($\psi =\chi \neq 0$).
  The composition of such solutions
give rise to 2-soliton and breather solutions.  These  are classified according to {\it
neutral-neutral} (2-solitons of the abelian  $A_{r-1}$ affine abelian Toda model ), {\it neutral-
charged} and  {\it charged-charged} solutions.   
The time delays were  also considered.

{\bf Acknowledgements} We are grateful to CNPq and Fapesp for  financial support.  One of us (JFG) thanks
Fapesp  and the organizers of the XXIII International Colloquium on Group Theoretical Methods in
Physics for financial support and kind hospitality.


\begin{thebibliography}{99}




\bibitem{lez-sav}  A. N. Leznov, M. V. Saveliev, Group Theoretical Methods for 
Integration of
Nonlinear Dynamical Systems, Progress in Physics, Vol. 15 (1992), Birkhauser
Verlag, Berlin; 
A. N. Leznov and M. V. Saveliev, Commun. Math. Phys. {\bf 89} (1983) 59


\bibitem{ime}
J.F. Gomes, F.E.M. da Silveira, G.M. Sotkov and A.H. Zimerman,
``Singular Non-Abelian Toda
Theories'', hepth 9810057,  in  ``Nonassociative
Algebras and its Applications'', Lec. Notes in Mathematics, Vol 211,p. 125-136
 Ed. R. Costa et. al., Marcel Dekker, (1999).
J.F. Gomes, E. P. Gueuvoghlanian, F.E.M. da Silveira, G.M. Sotkov and 
A.H. Zimerman,
``Singular Conformal, and Conformal Affine  Non-Abelian Toda Theories'', 
``M.V. Saveliev Memorial Volume'', Dubna (2000), Ed. A.N. Sissakian, p. 38-50,
 see also hepth/0002173.
 
\bibitem{lund} F. Lund, Ann. of Phys. {\bf 415} (1978) 251;\\
 F. Lund and T. Regge, Phys. Rev. {\bf D14} (1976) 1524
 
\bibitem{Aratyn} H. Aratyn, L.A. Ferreira, J.F. Gomes and A.H. Zimerman,
Phys. Lett {\bf B 254} (1991) 372


\bibitem{dual}
J.F. Gomes, E. P. Gueuvoghlanian ,F.E.M. da Silveira, G.M. Sotkov and 
A.H. Zimerman, `` T-duality of axial and vector dyonic integrable models, hepth/0007116

\bibitem{luis}
L.A. Ferreira, J.L. Miramontes and J.S. Guillen, 
J. Math. Phys. {\bf 38}, (1997), 882-901

\bibitem{Witten1}
E. Witten, Phys. Rev. Lett. {\bf 38} (1978) 121

\bibitem{emilio} J.F. Gomes, E.P. Gueuvoghlanian, G.M. Sotkov and A.H.
Zimerman, IFT-Unesp preprint, IFT-P-064-2000,
 ``Electrically Charged Topological Solitons''
 also in hep-th/0007169;  
 J.F. Gomes, E.P. Gueuvoghlanian, G.M. Sotkov and A.H.
Zimerman, IFT-Unesp preprint (in preparation) ``Dyonic Integrable Models''



\bibitem{plb}
 J.F. Gomes,  G.M. Sotkov and A.H. Zimerman, 
 Phys. Lett. {\bf 435B} (1998) 49, also in hepth/9803122,
  Ann. of Phys.{\bf 274}, (1999), 289-362, also in hepth/9803234
 
\bibitem{gervais-saveliev} 
J.-L. Gervais and M. V. Saveliev, Phys. Lett. {\bf 286B} (1992) 271


\bibitem{vertex}H. Aratyn, L.A. Ferreira, J.F. Gomes and A.H. Zimerman,
Supersymmetry and Integrable models (Lecture Notes in Physics Vol. 502), Ed. H.
Aratyn et. al. (Berlin, Springer), p. 197-210, (1998); J.
Phys. {\bf A31}, (1998) 9483-9492



\end{thebibliography}
\end{document}